\documentclass{article}
\usepackage{ijcai19}

\usepackage{times}
\usepackage{soul}
\usepackage{url}
\usepackage{amsmath}
\usepackage[
hidelinks,
colorlinks=true,      
linkcolor=black,       
citecolor=black,       
urlcolor=blue 
]{hyperref}
\usepackage[utf8]{inputenc}
\usepackage[small]{caption}
\usepackage{graphicx}
\usepackage[printonlyused]{acronym}
\usepackage{natbib}
\usepackage{array}

\title{Generative Simulations of The Solar Corona Evolution With Denoising Diffusion : Proof of Concept}

\author{
\large\textnormal{
\newline
Grégoire Francisco$^{1,2,3}$\footnote{Contact Author : gregoire.francisco@gmail.com}\and
Francesco Pio Ramunno$^{4,5}$\and
Manolis K. Georgoulis$^{6,7}$\and
João Fernandes$^8$\and  \\
Teresa Barata$^2$\And
Dario Del Moro$^1$\\
}
\affiliations
\small{\textit{
\newline
$^1$Department of Physics, University of Rome Tor Vergata, Rome, Italy\\
$^2$IA, Instituto De Astrofisica E Ciências Do Espaço, University of Coimbra, Coimbra, Portugal\\
$^3$Department of Physics, University of Rome La Sapienza, Rome, Italy\\
$^4$Institute for Data Science, University of Applied Sciences North Western Switzerland (FHNW), 5210 Windisch, Switzerland\\
$^5$Department of Computer Science, University of Geneva, 1211 Geneva, Switzerland\\
$^6$ Johns Hopkins University Applied Physics Laboratory Laurel, MD 20735, USA\\
$^7$ Research Center for Astronomy and Applied Mathematics of the Academy of Athens, 11527 Athens, Greece \\
$^8$CITEUC, Geophysical and Astronomical Observatory, University of Coimbra and Department of Mathematics, Coimbra, Portugal \\
}}
}

\begin{document}

\maketitle

\begin{abstract}
The solar magnetized corona is responsible for various manifestations with a space weather impact, such as flares, coronal mass ejections (CMEs) and, naturally, the solar wind. Modeling the corona's dynamics and evolution is therefore critical for improving our ability to predict space weather
In this work, we demonstrate that generative deep learning methods, such as Denoising Diffusion Probabilistic Models (DDPM), can be successfully applied to simulate future evolutions of the corona as observed in Extreme Ultraviolet (EUV) wavelengths. 
Our model takes a 12-hour video of an Active Region (AR) as input and simulate the potential evolution of the AR over the subsequent 12 hours, with a time-resolution of two hours. 
We propose a light UNet backbone architecture adapted to our problem by adding 1D temporal convolutions after each classical 2D spatial ones, and spatio-temporal attention in the bottleneck part. 
The model not only produce visually realistic outputs but also captures the inherent stochasticity of the system's evolution. 
Notably, the simulations enable the generation of reliable confidence intervals for key predictive metrics such as the EUV peak flux and fluence of the ARs, paving the way for probabilistic and interpretable space weather forecasting.
Future studies will focus on shorter forecasting horizons with increased spatial and temporal resolution, aiming at reducing the uncertainty of the simulations and providing practical applications for space weather forecasting.
The code used for this study is available at the following link: \href{https://github.com/gfrancisco20/video_diffusion}{https://github.com/gfrancisco20/video\_diffusion}.

\end{abstract}

\section{Introduction}

Key physical processes with a space weather impact occur within the solar corona.
One example are the magnetic reconnection episodes resulting in solar flares (\cite{Priest2002}).
Another one is the formation of coronal holes, known to be a determinant driver in the production of fast solar winds (\cite{Cranmer2002},\cite{Wang2024}).
Reliable modeling of the solar corona's evolution is thus crucial for enhancing space weather forecasting capabilities. 
To that end, a significant body of research focuses on the extrapolation of coronal magnetic fields from photospheric magnetograms.
To do so, many approaches leverage the fact that the dominant forces are magnetic and thereby model the magnetohydrodynamic (MHD) system as a magnetostatic one by neglecting all other forces (\cite{Wiegelmann2012}). 
However, such assumptions are less realistic at higher layers of the solar  atmosphere, and the highly non-linear nature of coronal MHD processes like flares (\cite{Shibata2011}), combined with the limitations of available boundary conditions, makes these models and other non-force-free ones computationally expensive and impractical for long-term forecasts. 
Moreover, the stochastic emergence of new magnetic flux from the solar interior adds further complexity to these models (\cite{Cheung2014}).

In this work, we explore the potential of generative deep learning methods to efficiently model such complex systems. 
Generative latent variable models have demonstrated their strong abilities in learning the mapping of high-dimensional data distributions (e.g., images or videos) onto latent probabilistic spaces, allowing for realistic sampling and extrapolation (\cite{Asperti2023}).
Among such methods \ac{VAE}s (\cite{Kingma2013}) offer computational efficiency but tend to struggle with high-frequencies and small-scale details. 
\ac{GAN}s (\cite{Goodfellow2014}) partially address these shortcomings by producing higher-quality samples, but are difficult to train and often suffer from mode collapse, leading to poor diversity in generated data.
Recently, \ac{DDPM}s have been shown to consistently outperform both \ac{VAE}s and \ac{GAN}s in terms of sample quality and diversity across several applications, including image generation, super-resolution, and image-to-image translation (\cite{Ho2020}). 
The stochastic generation process of diffusion models makes them particularly suitable for capturing the intrinsic stochastic nature of non-linear physical systems (\cite{Sohl-Dickstein2015}). 
Recent studies showed promising applications in the field of space weather.
\cite{Ramunno2023} successfully used \ac{DDPM}s to model a latent distribution of solar EUV observations, and generate realistic new synthetic samples.
\cite{Ramunno2024} showed that \ac{DDPM}s can forecast photospheric magnetogram 24 hours before flares, outperforming persistence models both in computer science and physical performance indicators.
In this study, we aim to extend the application of \ac{DDPM}s to video-to-video forecasting of \ac{SDO}/\ac{AIA} observations at the 94\AA\ wavelength. 
The 94\AA\ channel is particularly important for studying solar flares, as it is most sensitive to plasma temperatures around 6.3 million K, which are closely associated with flare activity \cite{Boerner2012}. 
Previous work has shown that emissions in the 94\AA\ and 131\AA\ channels are strongly correlated with the peak flux of \ac{SXR} during flares, and can provide reliable proxy for flare magnitude in \ac{SXR} (\cite{Kiera2022}). 
Reliable probabilistic simulations of the corona evolution in these wavelength could thus aid in both flare magnitude estimation and flare timing prediction, the latter being one of the hardest challenges in space weather forecasting (\cite{Boucheron2015}).
The 94\AA\ wavelength and a 2-hour temporal resolution used in this study serve as a proof of concept to illustrate the potential of the approach for space weather forecasting and solar physic studies. 
Future works will focus on extending this approach to additional wavelengths and higher temporal resolution to reduce the resulting simulations uncertainty and provide practical space weather applications.
To reduce computational complexity while preserving the physical fidelity of active region (AR) dynamics, we limit our analysis to ARs within ±230 arcsec from the solar disk center, thus minimizing the effects of projection and solar rotation.
This enables to highlight the model's ability to learn the intrinsic physical dynamics of ARs.
The paper is organized as follows: Section \ref{sec:model} describes the model, starting with an introduction to key \ac{DDPM} concepts followed by the description of our light UNet backbone to learn temporal patterns from videos. 
Section \ref{sec:dataset} presents the dataset used for training and evaluation. 
Section \ref{sec:results} discusses the performances, while Section \ref{sec:discussion} concludes with a discussion of our findings and future works.

\section{Model}\label{sec:model}

\subsection{Background}\label{sec:model:background}

\subsubsection{Denoising Diffusion Probabilistic Model (DDPM)}

A diffusion model (\cite{Sohl-Dickstein2015}) $p_{\theta}$ is defined as a reverse process modelling a variable $X_{0}$ as the $T-th$ state of a Markov Chain $(X_{T-t})_{t=0}^{T}$ of transitions defined such as : 
\begin{equation}
\begin{aligned}
    p_{\theta}(X_{T}) &:=  \mathcal{N}(X_{T} ; 0 , I)  \\ 
    p_{\theta}(X_{t-1}|X{t}) &:=  \mathcal{N}(X_{t-1} ; \mu_{\theta}(X_{t},t) , \Sigma_{\theta}(X_{t},t)) 
\end{aligned}\label{eq:trans_rev}
\end{equation}
In the \ac{DDPM} case, the unknown transitions $p_{\theta}(X_{t-1}|X{t})$ are modeled as the reverse of a forward diffusion of Gaussian transitions.
This forward process $q$ gradually adds noise to $X_0$, following a scheduled variance $(\beta_{t})_{t=1}^{T}$, that results in equilibrium at $T$, thus reaching the pure noise state $X_{T}$ :
\begin{equation}
\begin{aligned}
    q(X_{t}|X{t-1}) &:=  \mathcal{N}(X_{t} ; \sqrt{1-\beta_{t}} , \beta_{t}I)
\end{aligned}\label{eq:trans_fwd}
\end{equation}
This diffusion process can be resolved at any arbitrary timestep t, defining $\alpha_{t}:=1-\beta_{t}$ and $\hat{\alpha}_{t}:=\Pi_{i=0}^{t}\alpha_{i}$, with :
\begin{equation}
\begin{aligned}
    q(X_{t}|X_{0}) &:=  \mathcal{N}(X_{t} ; \sqrt{\hat{\alpha}_{t}}X_{0} , (1-\hat{\alpha}_{t})I)
\end{aligned}\label{eq:Qt}
\end{equation}
An autoencoder $\epsilon_{\theta}$ can then be trained to learn predicting a random noise $\epsilon \sim \mathcal{N}(0,I)$, added during the diffusion process at timesteps $t$ following the Equation \ref{eq:Qt}, i.e. $x_{t} := \sqrt{\hat{\alpha}_{t}}x_{0}+\sqrt{1-\hat{\alpha}_{t}}\epsilon$. 
Specifically, the training is performed by minimising a distance - typically L2 - between the actual noise and its autoencoder's estimations $\epsilon_{\theta}(x_{t},t)$:
\begin{equation}
\begin{aligned}
    min & \lVert \epsilon -  \epsilon_{\theta}(x_{t},t) \rVert_2 \ , \ \forall t \in [\![1,T ]\!]
\end{aligned}\label{eq:training}
\end{equation}
Once $\epsilon_{\theta}$ trained, the predicted noise can be used to approximate $\mu_{\theta}(X_{t},t)$, the mean of the reverse process, as derived in \cite{Ho2020}:
\begin{equation}
\begin{aligned}
\frac{1}{\sqrt{\hat{\alpha}_{t}}}(x_{t}-\frac{\beta_{t}}{\sqrt{1-\hat{\alpha}_{t}}} \epsilon_{\theta}(x_{t},t))
\end{aligned}\label{eq:inference}
\end{equation}
The variance $\Sigma_{\theta}(X_{t},t)$ can be fixed to $\beta_{t}$ to match the forward process's variance schedule.
Using these values in Equation \ref{eq:trans_rev}, new samples $x_{0}$ can be generated by iteratively reversing the diffusion process, starting from pure random noises $x_{T}  \sim \mathcal{N}(0,I)$.
The computational cost of this sampling procedure depends on the noise schedule $(\beta_{t})_{t=1}^{T}$, which must be chosen so that the Signal-To-Noise-Ratio (SNR) becomes null in T, i.e. $SNR_{T} = \frac{\hat{\alpha}_{T}}{1-\hat{\alpha}_{T}} \approx 0$, in order for $X_{T}$ to be pure noise.
In the case of the linear noise schedule proposed by \cite{Ho2020}, such condition start being satisfied from $T \approx 1000$ as $SNR_{1000} \approx 4e-05$.
\subsubsection{Conditional Diffusion}
Finally, all the equations of this section can be extended to handle conditional diffusion by marginalizing the probability distributions over any conditioning variable $c$.
In our case the conditions $c$ represents the previous 12 hours of input data.
These conditions are projected by the autoencoder into a continuous latent variable space, allowing the model to generalize to unseen conditions during training and generating predictions for new video sequences.
Specifically, for any $c$ representing 12 hours of an active region at 2 hours resolution, our models learn to simulate possible realisations of the next 12 hours $X_{0}$, with a latent space approximating the conditional transitions:
\begin{equation}
\begin{aligned}
    p_{\theta}(X_{t-1}|X{t},c) &:=  \mathcal{N}(X_{t-1} ; \mu_{\theta}(X_{t},t,c) , \Sigma_{\theta}(X_{t},t)) 
\end{aligned}
\end{equation}

\subsection{Light Video UNet}\label{sec:model:unet}

For the autoencoder, we use a modified version of the Palette's UNet architecture (\cite{Saharia2021}) also used by \cite{Ramunno2024} for magnetogram-to-magnetogram forecasting.
In our case, to adapt the model to videos and improve its ability to capture temporal patterns, we add temporal 1D-convolutions after each spatial 2D-convolutions.
This approach provides spatio-temporal feature learning abilities similar to 3D-convolutions but with lower computational complexity as the complexity of a 2D+1D convolutions block is $O(n^{2}+n)$ against $O(n^{3})$ for 3D-convolutions, with $n$, the size of the kernels.
For more complex patterns, we incorporate spatio-temporal self-attention (\cite{Vaswani2017}) blocks between each convolution block in the bottleneck.
As the feature-maps size is smallest in the bottleneck - 16x16 in our model -, this placement allows for attention scores computation at the smallest possible complexity.

\section{Dataset}\label{sec:dataset}

For the of \ac{AIA} \ac{EUV} images we start from the SDO-2H-ML dataset that we defined in \cite{Francisco2024}.
Specifically, the images are aligned, exposure normalised, instrument degradation corrected, downsampled at 1024 by 1024 pixels and  available at a 2 hour resolution.
Their pixel values are logarithmically scaled and depth-downsampled to 8-bit, which allow to preserve most of the original pixel distribution with marginal information loss, while making the dataset more compact.  
In this work we use the resulting images before the original JPEG compression used in \cite{Francisco2024}.
As we seek to test the model's ability to learn and simulate \ac{AR}s' dynamics, while limiting the impact of the solar rotation and projection effects, we focus on crops of 614 by 614 arcsec centered around known SHARPs (\cite{Bobra2014}) that are within ±230 arcsec from the solar disk center.
The important size of our crops, relatively to original SHARPs bounding boxes, is meant to ensure capturing all the coronal loops that can be associated to a given \ac{AR}.
On the other end, this approach imposes us a strict chronological split between training and test data, as some of our crops may partially overlap over each other.
The resulting crops are further downsampled to 128 by 128 pixels for computational reasons.
Samples are created at a 2 hour cadence when possible by pairing the crops of a given \ac{AR} within [-10h,0h] for the input video, with the crops within [+2h, +12h] for the targets, resulting in pairs of 6 frames videos.
While our target is only the 94\AA \ wavelenght, we add the 193\AA \ and 211\AA \ wavelength in the input to benefit from more information on coronal dynamics at different temperatures.
We train on the resulting samples between 2010-05 and 2022-03, which amount to a total of 34000 paired videos.
We test on samples spanning from 2022-05 to 2023-04 that are buffered by at least one month from the training data.
Due to our spatial and temporal constraints, our test samples do not exhibit any X-flares, and only 41 samples exhibit M-flares.
To build a a balanced test, we randomly select 41 C-flares, and 41 samples without any flare above the C-threshold, such sample will be referred a quiet samples.
This results in a small test of 123 samples that allow to estimate our probabilistic model performances at a moderate cost.

\section{Results}\label{sec:results}

\subsection{Computer vision metrics}\label{sec:results:mtc_cv}

To evaluate the quality of the generated videos, we propose using a set of classical computer vision metrics: Peak Signal-to-Noise Ratio (\ac{PSNR}), Structural Similarity Index (\ac{SSIM}), and Learned Perceptual Image Patch Similarity (\ac{LPIPS}). Although these metrics are typically designed for image comparison, we apply them frame-wise and calculate their averages to derive an overall score for the video.
The \ac{PSNR} compares the pixel-wise mean squared error to the maximum possible value of the signal, normalizing the distortion of the generated videos by the potential dynamic range. In the case of Extreme Ultraviolet (EUV) observations, the signal can vary across several orders of magnitude, corresponding to different physical processes. To more accurately assess the model's ability to preserve the underlying physics, we apply this metric to the descaled pixel distribution, reverting it to its original dynamic range. Consequently, the resulting \ac{PSNR} scores appear much higher than those typically computed over standard 8-bit images.
For context, we refer to \cite{Chervyakov2023}, which derived a threshold formula indicating that a \ac{PSNR} value above $Q = 5*log(max_{signal-value}) / log(2)$ suggests low distortion and high-quality images. With our saturation value of 6099 DN/s for the 94 Å EUV wavelength, this roughly corresponds to $Q\approx63$.
However, pixel-wise metrics like \ac{PSNR} may not be the most relevant in our case. Given the non-deterministic nature of our problem, it is more suitable to compare the presence of similar structures in the images with slightly relaxed conditions regarding their pixel-wise shape and locations. To this end, the Structural Similarity Index (\ac{SSIM}) is particularly relevant, as it evaluates a combination of indicators derived from the first and second-order moments of the pixel distribution over a small window - in our case, 11 by 11 pixels. Images are considered to have good similarity for \ac{SSIM} values above 0.7. Similar to \ac{PSNR}, we compute \ac{SSIM} using the original (descaled) pixel dynamic range.
Finally, for a more intuitive interpretation of perceptual similarity between the images, we use \ac{LPIPS} (\cite{Zhang2018}) directly on the 8-bit generated videos, with values below 0.1 indicating good similarity.

\subsection{Space weather related metrics}\label{sec:results:mtc_phy}

To further evaluate the model's ability to capture key physical quantities, we complement the previous metrics with physical metrics derived from the \ac{EUV} flux measured within the frames. 
While an estimate of the actual EUV flux can be obtained using the corresponding instrument's response function (\cite{Boerner2012}, \cite{Barnes2020}), we simplify our metrics by using the DN/s values as proxies for the flux. 
This simplification is particularly relevant, as \cite{Kiera2022} demonstrated that DN/s values from \ac{AIA})observations can accurately approximate Soft X-Ray (SXR) flux, which is of particular interest in space weather forecasting.
Consequently, for each frame, our proxy for the corresponding region's flux is the sum of the pixel DN/s values. 
From the resulting flux time series, we focus on two essential derived quantities: the \ac{MPF} and the fluence.
The \ac{MPF} is defined as the maximum value of the flux within the entire time series. 
It is particularly relevant as it provides an estimate of the maximum activity expected during the forecasted time window. 
This metric is well suited to the stochastic nature of the problem, as we anticipate some uncertainty regarding the precise timing of flare occurrences, especially when forecasting over several hours. 
To complement the \ac{MPF}, we introduce the \ac{T2PF}, which indicates the time remaining before reaching the \ac{MPF} and poses a particularly challenging problem in flare forecasting (\cite{Boucheron2015}).
The fluence is defined as the integral of the flux over the considered period, approximated by:
\begin{equation}
    fluence = \Sigma_{t} (\Delta_{t} * flux_{t})
\end{equation}
where $\Delta_{t}$ corresponds to our 2-hour resolution converted to seconds. 
Due to our relatively low 2-hour time resolution, the resulting \ac{MPF} and fluence values are not meaningful approximations of the actual physical values for the corresponding time windows. 
Therefore, these metrics are not yet intended for practical applications but serve as indicators of the model's ability to capture key physical quantities at a given resolution. Specifically, we focus on the model's \ac{MAPE} in \ac{MPF} and fluence, as well as the \ac{MAE} in \ac{T2PF}.
Finally, we evaluate the model's ability to generate realistic trajectories of the flux time series by computing the \ac{DTW} Euclidean distance (\cite{Piersol1981}) between the generated time series and the ground truth. 
The \ac{DTW} metric allows us to estimate the similarity between two time series while correcting for the impact of small delays or variations in speed. This feature enables better comparisons when the sequences may be out of phase, as is the case here. 
To enhance interpretability, we focus on the \ac{DTW} distance normalized by the Euclidean norm of the observed time series.

\subsection{Performances}\label{sec:results:perf}

The performance metrics of the models are presented in Table \ref{tab:perf}. 
For each sample 20 simulations are generated, and the final score per sample is calculated as the average score over all the simulations.
The good scores in the first three columns, corresponding to the computer science metrics, suggest that the results are perceptually realistic. 
Both \ac{PSNR} and \ac{LPIPS} show slight degradation during periods when stronger flare events occurred within the forecasting window, indicating greater uncertainty at the pixel level and a lower perceptual similarity, respectively. 
On the other hand, the \ac{SSIM} improves, suggesting that the model performs better at a structural level, as the metric compare the images similarity over small local windows of a few pixels.
This apparent contradiction between the \ac{PSNR} and \ac{SSIM} may stem from the model's ability to predict accurately brightenings, such as flares, while exhibiting some uncertainty about their precise pixel-wise shape and location. 
During higher solar activity, such brightenings constitute a larger portion of the overall signal, leading to the increased \ac{SSIM}.
The last four columns indicate that the model successfully captures essential physical properties. 
For instance, the mean absolute percentage error for both MPF and fluence is approximately 40\%. 
When breaking down the performance by activity levels, the errors remain within the same order of magnitude of the observed values. 
While promising for practical space weather applications, it is noted that the low temporal resolution of this study smooth the flux time series, simplifying the task and rendering the model not suitable for operational space weather forecasting, as most flares' peak do not appear in the training and evaluation data.

\begin{table*}[h!]
\centering
\begin{tabular*}{\textwidth}{@{\extracolsep{\fill}} l||c|c|c||c|c|c|c c@{}}
\hline\hline
\small{Models} & \small{PSNR\ $\uparrow$} & \small{SSIM\ $\uparrow$} & \small{LPIPS \ $\downarrow$} & \small{Flux-}\footnotesize{DTW} & \small{MPF} \scriptsize{(MAPE)} & \small{Fluence} \scriptsize{(MAPE)} & \small{T2PF} \scriptsize{(MAE)} \\ \hline \vspace{-1.9ex} & & & & & & &  \\ 
\small{DDPM} & 73 & 0.77 & 0.09 & 1.00  & 42\% & 41\% & 4.84H  \\ 
            & ± 2.4 & ± 0.13 & ± 0.02 & ± 0.57 & ± 25 & ± 24 & ± 2.53  \\ \hline \vspace{-1.9ex} & & & & & & & \\  
\small{DDPM (quiet)} & 79 & 0.67 & 0.07 & 0.99  & 38\% & 40\% & 4.93H \\ 
            & ± 3.1 & ± 0.15 & ± 0.03 & ± 0.64 & ± 0.27 & ± 0.26 & ± 2.65 \\ \hline \vspace{-1.9ex} & & & & & & & \\ 
\small{DDPM (C)} & 72 & 0.80 & 0.08 & 0.93  & 38\% & 38\% & 5.31H \\
            & ± 2.3 & ± 0.13 & ± 0.02 & ± 0.55 & ± 0.23 & ± 0.23 & ± 2.64 \\ \hline \vspace{-1.9ex} & & & & & & & \\ 
\small{DDPM (M)} & 67 & 0.84 & 0.11 & 1.09  & 49\% & 44\% & 4.29H \\
            & ± 1.9 & ± 0.11 & ± 0.03 & ± 0.53 & ± 0.24 & ± 0.23 & ± 2.31 \\ 

\hline
\end{tabular*}
\caption{Models Performances. 
The first row corresponds to the model's performances over the all  test set presented in section \ref{sec:dataset}.
The three following row indicates the performances of the model on the restriction of samples which exhibited the following activity during the forecasted window : quiet for the second row, at least one C-class flare for the third row, at least one M-class flare for the fourth row.
The first three columns are computer science metrics (section \ref{sec:results:mtc_cv}) where $\uparrow$ indicates that higher scores are better, and $\downarrow$ that lower scores are better.
The last columns correspond display the error in the physical metrics presented in section \ref{sec:results:mtc_phy}.
For every sample 20 simulations are generated. 
The first rows of the cells indicate the average score over all the simulations.
The second rows of the cells (± symbol) indicate the average standard deviation of the sample in the corresponding scores.
}\label{tab:perf}
\end{table*}

\begin{figure*}[h!]
    \centering
    \textbf{Reliability Diagrams}\par\medskip
    \hspace{-5ex}
    \includegraphics[width=0.98\textwidth]{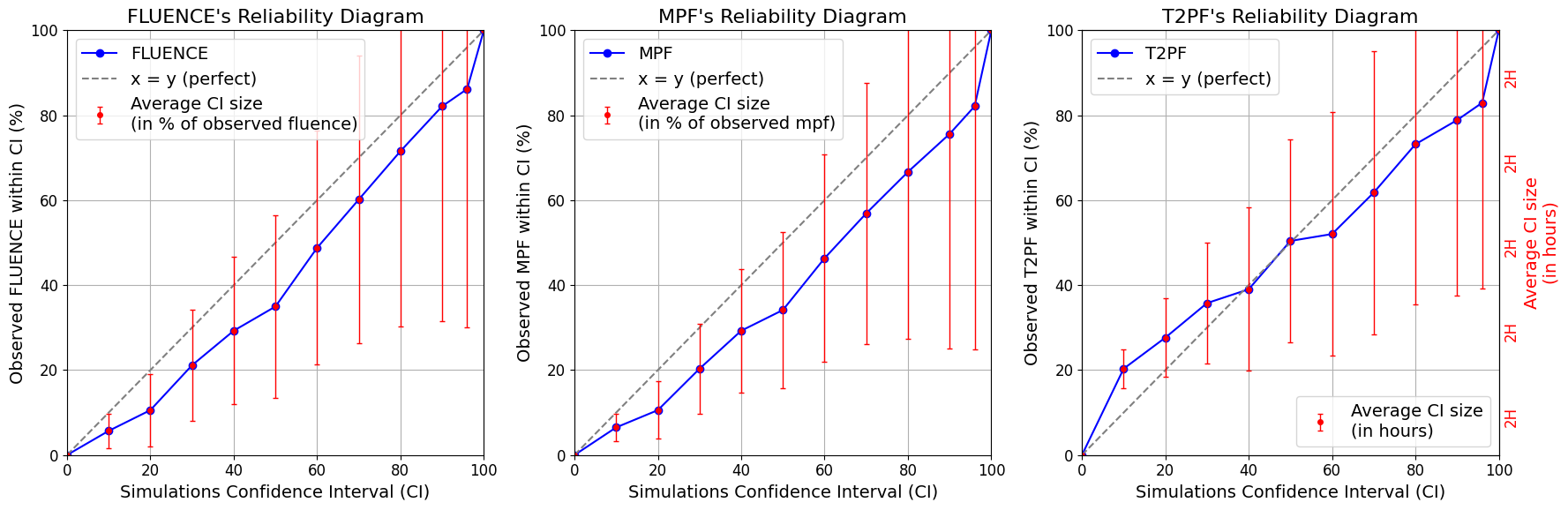}
    \caption{
    Reliability diagrams. The X-axis represents the Confidence Intervals (CIs) derived from the simulations. The left Y-axis represents the frequency at which actual observations fall into the derived CIs, corresponding to the blue curve. The red bars represent the size of the CIs, expressed as a percentage of the observed value for fluence and \ac{MPF}, with one tile on the Y-axis grid representing 20\% uncertainty. For the \ac{T2PF}, the red bar scale (right Y-axis) is in hours, with each tile representing 2H of uncertainty. A perfectly reliable model has a blue curve aligned with the diagonal and provides minimal uncertainty for a given confidence interval, represented by smaller red bars.
    }
    \label{fig:diagrams}
\end{figure*}

From the 20 simulations generated for each sample, estimates of the fluence, \ac{MPF}, and \ac{T2PF} distributions were derived. 
Their reliability are assessed with the reliability diagrams shown in Figure \ref{fig:diagrams}.
The blue curves depict how frequently the ground truth falls within the corresponding \ac{CI}. 
For a perfectly reliable probabilistic model, these curves would align with the diagonal ($y = x$), meaning that observations would fall within the p\%-CIs p\% of the time for any p in the range [0,100]. 
The red bars indicate the size of the CIs, reflecting the level of uncertainty at a given confidence level. 
For practical purposes, smaller uncertainties are preferable, with the ideal model having CIs as small as possible.

For all three metrics, the generated probability distributions appear reliable, with the blue curves aligning with the diagonal at a few point of percentage. 
This suggests that the model effectively captures the stochastic nature of coronal evolution, which arises from the system's inherent non-linearity and the stochastic emergence of new flux. 
The model's resulting stochasticity is then further emphasized by the limited input information, such as the low spatial and temporal resolutions. 
As a result, the uncertainty of the generated simulation is quiet large, with 96\%-CIs reaching approximately ±55\% of the actual observation for the fluence and the \ac{MPF}, and ±4H for the \ac{T2PF}.

\subsection{Prediction examples}\label{sec:results:pred}

\begin{figure*}[h!]
    \centering
    \textbf{Predictions example on AR 8195 the 2022-05-04 at 20:00}\par\medskip
    \hspace{-5ex}
    \includegraphics[width=0.98\textwidth]{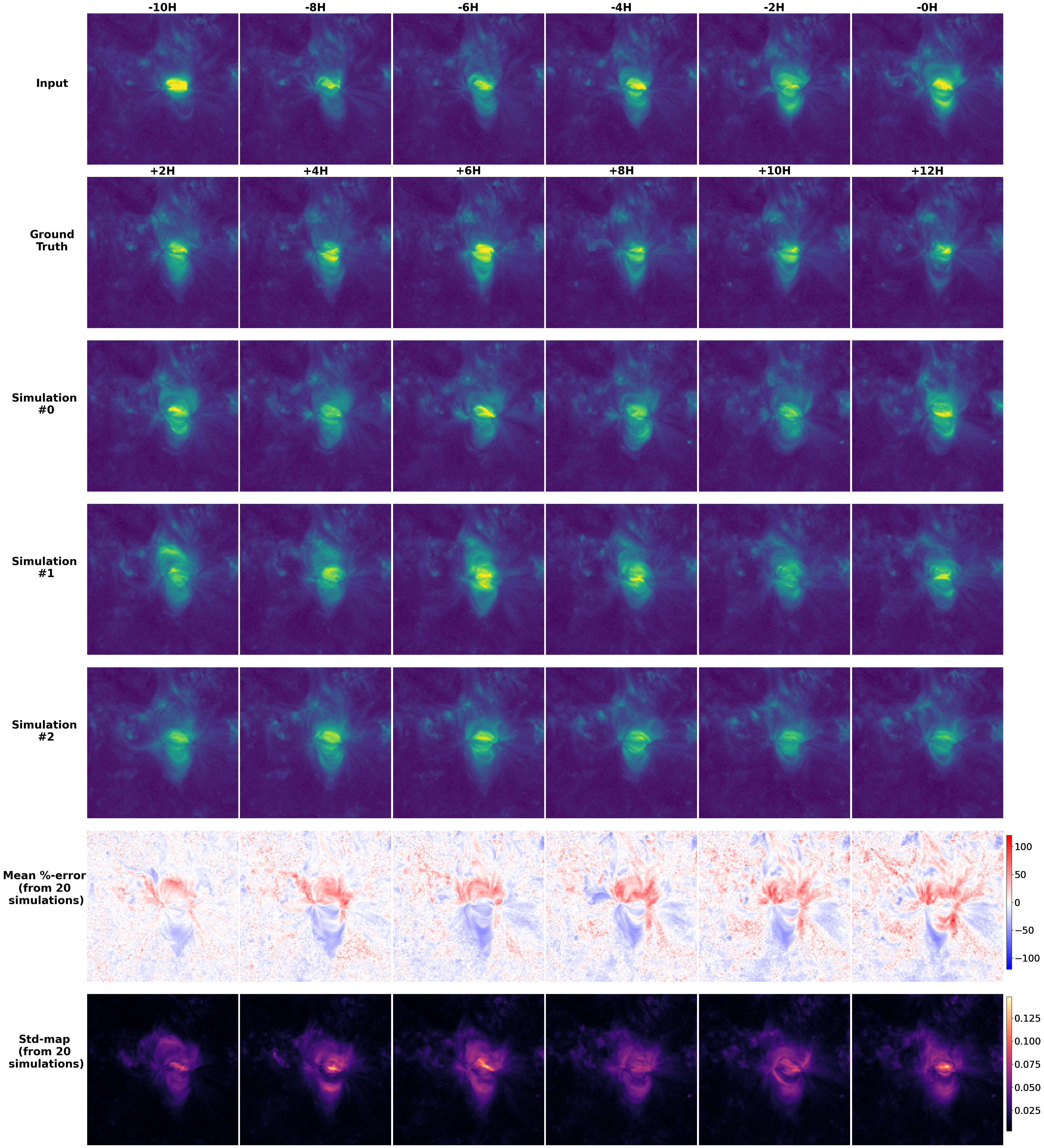}
    \caption{
    Predictions for AR 8195 on 2022-05-04 at 20:00. 
    The AR exhibited a series of numerous C-class and M-class flares during the 12 hours prior to the prediction time, as well as during subsequent 12 hours. 
    Remnants of these flares can be observed as brightenings in the frames, both in the first row (previous 12 hours) and in the second row (next 12 hours ground truth). 
    Similar brightening can be observed in the three simulation results exhibited from the third to fifth row.
    The sixth row presents the pixel-wise percentage deviation from the ground truth, averaged over 20 simulations. 
    The final row displays standard deviation maps computed over the same 20 simulations. Animations links : \href{https://github.com/gfrancisco20/video_diffusion/blob/master/paper_simulations/8195_20220504_2000_simu_0.gif}{Simulation \#0}, \href{https://github.com/gfrancisco20/video_diffusion/blob/master/paper_simulations/8195_20220504_2000_simu_1.gif}{Simulation \#1}, \href{https://github.com/gfrancisco20/video_diffusion/blob/master/paper_simulations/8195_20220504_2000_simu_2.gif}{Simulation \#2}.
}\label{fig:pred_8195}
\end{figure*}

\begin{figure*}[h!]
    \centering
    \textbf{Predictions example on AR 8977 the 2023-01-18 at 06:00}\par\medskip
    \hspace{-5ex}
    \includegraphics[width=0.98\textwidth]{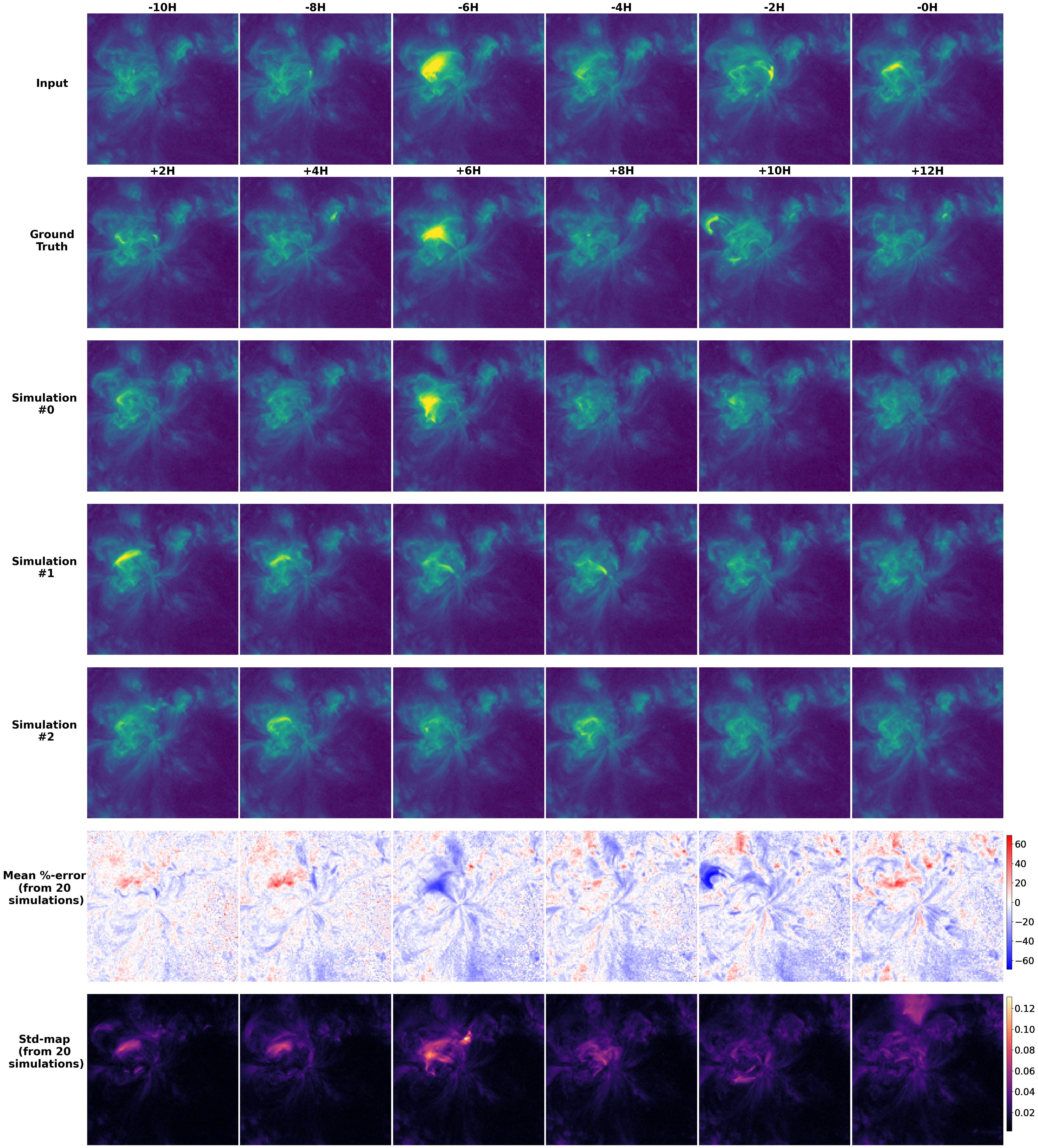}
    \caption{
    Predictions for AR 8977 on 2023-01-18 at 06:00. The -6H input frame shows the remnants of an M1.8 flare that occurred approximately 6.5 hours prior to the time of prediction. In the +6H ground truth frame, which represents the forecasted target, we observe the remnants of another M1.8 flare that concluded on 2023-01-18 at 10:52. A similar brightening is visible in the +6H frame of Simulation 0. The mean percentage pixel-wise error, averaged over 20 simulations, indicates that errors are larger during this event.
    The standard deviation map shows the pixel-wise standard deviation across 20 simulations, providing insight into the areas where the model is most uncertain. This highlights regions at higher risk for extreme events, such as the flare remnants seen in the +6H frame. Animations links : \href{https://github.com/gfrancisco20/video_diffusion/blob/master/paper_simulations/8977_20230118_0600_simu_0.gif}{Simulation \#0}, \href{https://github.com/gfrancisco20/video_diffusion/blob/master/paper_simulations/8977_20230118_0600_simu_1.gif}{Simulation \#1}, \href{https://github.com/gfrancisco20/video_diffusion/blob/master/paper_simulations/8977_20230118_0600_simu_2.gif}{Simulation \#2}.
}
    \label{fig:pred_8977}
\end{figure*}

\begin{figure*}[h!]
    \centering
    \textbf{Predictions example on AR 9188 the 2023-03-08 at 22:00}\par\medskip
    \hspace{-5ex}
    \includegraphics[width=0.98\textwidth]{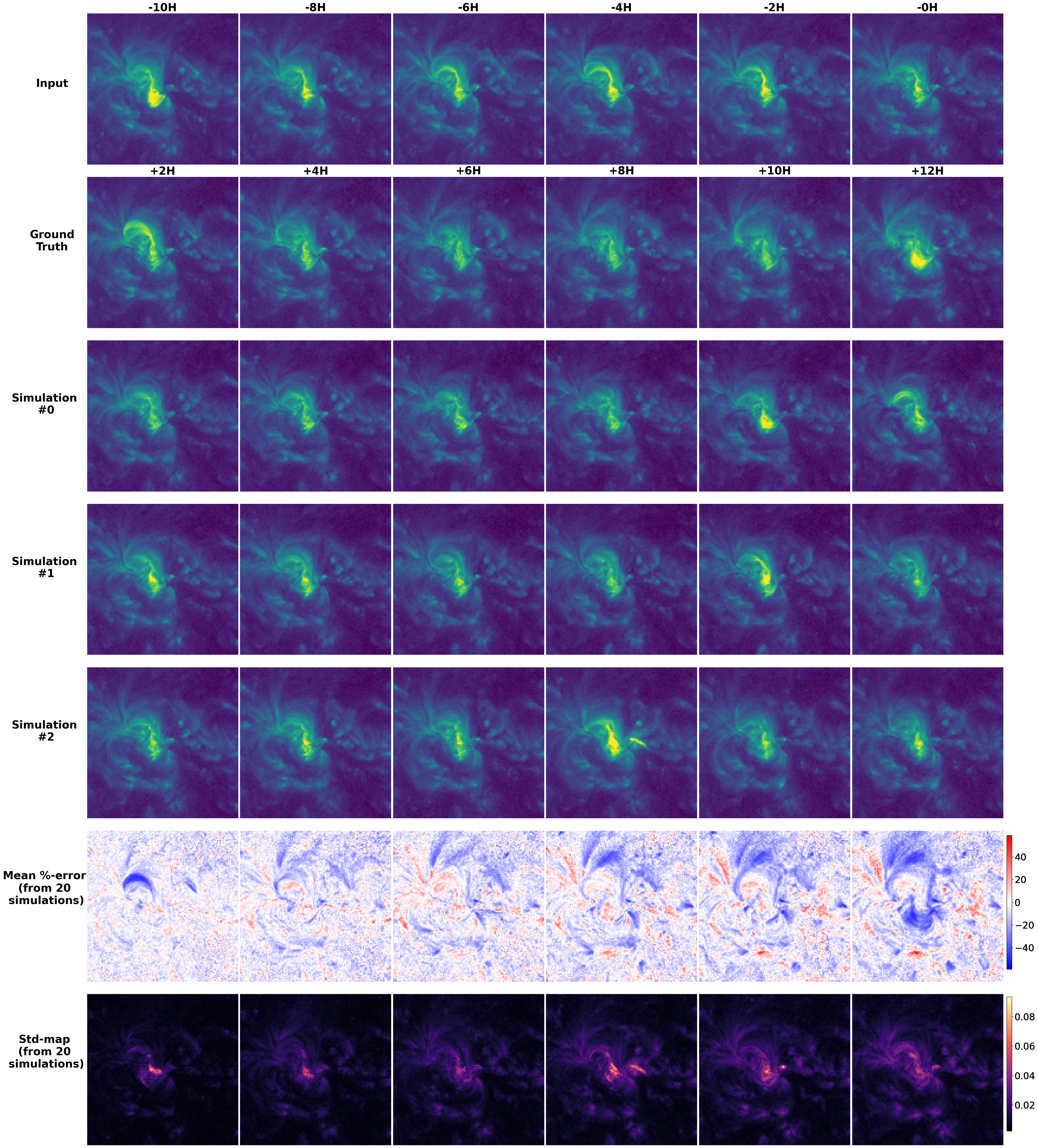}
    \caption{
    Predictions for AR 9188 on 2023-03-08 at 22:00. The AR exhibited a series of C-class and M-class flares during the [-10H, +12H] time window around the forecasting time. The -10H input frame shows strong remnants of a C flare that ended 30 minutes before the frame's timestamp. The +2H ground truth frame displays the fading remnants of an M1.3 flare, which ended about 1 hour and 10 minutes earlier. The +12H frame exhibits strong remnants of a C3.1 flare ending 40 minutes before. Simulations 0 and 1 display a similar brightening in the +10H frame, while Simulation 2 shows one in the +6H frame. The mean percentage error indicates that overall activity is mostly underestimated, particularly around the upper loop where the M-class flare occurred, showing difficulty in predicting this flare and its aftermath. The standard deviation map shows increased uncertainty in the lower part of the sigmoid, where the C3.1 flare occurs between the +10H and +12H frames, suggesting that the model forecasted the possibility of similar events between the +6H and +12H frames, where the standard deviations are higher. Animations links : \href{https://github.com/gfrancisco20/video_diffusion/blob/master/paper_simulations/9188_20230308_2200_simu_0.gif}{Simulation \#0}, \href{https://github.com/gfrancisco20/video_diffusion/blob/master/paper_simulations/9188_20230308_2200_simu_1.gif}{Simulation \#1}, \href{https://github.com/gfrancisco20/video_diffusion/blob/master/paper_simulations/9188_20230308_2200_simu_2.gif}{Simulation \#2}.
    }
    \label{fig:pred_9188}
\end{figure*}

To illustrate the diversity of the model's predictions, we present three simulations for each of the following cases: 
\begin{itemize} 
    \item AR 8195 on 2022-05-04 at 20:00, shown in Figure \ref{fig:pred_8195}
    \item AR 8977 on 2023-01-18 at 06:00, shown in Figure \ref{fig:pred_8977}
    \item AR 9188 on 2023-03-08 at 22:00, shown in Figure \ref{fig:pred_9188} 
\end{itemize} 
For each figure, the first row displays the sequence of frames from the 12 hours preceding the forecasting date. 
The second row corresponds to the actual frames for the next 12 hours, which the model aims to simulate. 
The third to fifth rows present three distinct simulations generated by the model. 
The sixth row shows the simulations' percentage deviation from the ground truth, averaged over 20 distinct simulations. 
The seventh and final row displays standard deviation maps, computed as the pixel-wise standard deviation across the same 20 simulations, highlighting the regions of the images where uncertainty is highest.

Comparing the mean percentage error across the three examples, the deviation from the ground truth appears balanced in the case of AR 8195 (Figure \ref{fig:pred_8195}), while the simulations for AR 8977 (Figure \ref{fig:pred_8977}) and AR 9188 (Figure \ref{fig:pred_9188}) displays in average an undercasting tendency.
The standard deviation maps highlight regions that are estimated to be at higher risk for extreme events. 
They mostly coincide with the areas and timesteps exhibiting the most intense activity in the ground truth. 
This suggests that, while the model does not assign a high probability to extreme events (indicated by the negative mean deviation), it captures their possibility with uncertainty around their precise timing and exact configuration.

The case of AR 8977 (Figure \ref{fig:pred_8977}) is particularly illustrative. 
An M1.8 flare began at 10:21 and ended at 10:52, 1 hour and 8 minutes before the +6H frame, where the ground truth exhibits strong remnants of the flare.
A very similar brightening is observed in the +6H frame of Simulation 0 but not in the other two simulations. 
Across the 20 generated simulations, this brightening appears twice at the same frame, resulting in an estimated probability of occurrence around 10\% at that specific time step. 
As a result, the standard deviation map is brightest for that frame and in the region of the flare.

Similar observations can be made for AR 9188 (Figure \ref{fig:pred_9188}), which exhibits a bright flare remnant in the +12H frame. 
In this case, the model generates similar brightening in the +10H frame for Simulations 0 and 1, and in the +8H frame for Simulation 3. 
This suggests that the model is relatively confident about the occurrence of such events within the final hours of the forecasted time window but remains uncertain about the precise timing. 
This uncertainty is also highlighted by the bright standard deviation maps for the +8H to +12H frames.

Finally, AR 8195 is marked by a series of several C-class and M-class flares during both the forecasted time window and the 12 previous hours.
This high level of activity results in complex coronal loop dynamics and configurations, which appear well understood by the model. 
Indeed, the model generates realistic and diverse possible evolutions for the active region, illustrating again the strength of the \ac{DDPM} in modeling the stochastic aspects of the problem.

\section{Discussion \& Conclusions}\label{sec:discussion}

Despite the constraints of using a low temporal resolution (2 hours) and focusing only on \ac{AR}s at the solar disk center, our current approach demonstrate promising potential.
In particular, our findings suggest that \ac{DDPM}s can be used to efficiently generate realistic evolutions of the solar corona as observed in \ac{EUV}. 
Our probabilistic video forecasting model appears well-calibrated and reliable for deriving key space weather indicators.
 As seen in our prediction examples, the model is able to simulate complex MHD dynamics and forecast brightening events with time steps closely aligned to their actual occurrences.
These encouraging outcomes suggest such models could represent a major milestone in the fields of Solar Physics and Space Weather.
While current space weather forecasting abilities are sometimes compared to earth weather forecasting abilities of 50 years ago, these new methods could significantly reduce that gap, by providing computationally efficient simulation tools to visualize the possible trajectories of the system's evolution.
Major limitations of this proof of concept must however be acknowledged.
The low temporal resolution of 2 hours, statistically excludes most actual flare peaks from the training and evaluation data, making the resulting model not suited to model the occurrence of such event.
The temporal and spatial restrictions of this work were intended solely to validate the concept with limited computational resources. 
A time resolution of 15 to 10 minutes could already allow to capture much better extreme transient fluctuations in variables of interest like the flux, resulting in improved flare predictions with reliable likelihood estimates and interpretable simulations of their occurrence.
At a time resolution of 2 minutes or less, a DDPM might further model the physical dynamics of the corona at a much finer scale, potentially resulting in reduced uncertainty and promising application for solar physics studies.
For instance, such models could be used to estimate the intrinsic stochasticity of flare occurrences given current instrument limitations in spatial and temporal resolution.
Moving forward, future work could thus focus on generating forecasts for 6-hour periods at a 10-minute resolution and for 1-hour periods at a 2-minute resolution. 
Additionally, increasing the spatial resolution to 2.4 arcsec/pixel might also be worth exploring, while training the model at any longitudes for full-disk coverage will be indispensable for operational applications.

\newpage
\textbf{\Large{Acknowledgements}}
\newline
\newline
\begingroup
\linespread{1}
\small
This research is part of the SWATNet project which is funded by the European Union’s Horizon 2020 research and innovation program under the Marie Sklodowska-Curie Grant Agreement No \href{https://doi.org/10.3030/955620}{955620}. 

This study was also produced within the IA and the CITEUC.
IA is supported by Fundação para a Ciência e a Tecnologia (FCT, Portugal) through the research grants \href{http://doi.org/10.54499/UIDB/04434/2020}{UIDB/04434/2020} and \href{http://doi.org/10.54499/UIDP/04434/2020}{UIDP/04434/2020}.
CITEUC is funded by National Funds through FCT - project UIDP+UIDB/00611/2019.

Project partially funded under the National Recovery and Resilience Plan (PNRR), Missione 4 “Istruzione e Ricerca” – Componente C2 – Investimento 1.1, “Fondo per il Programma Nazionale di Ricerca e Progetti di Rilevante Interesse Nazionale (PRIN)” – Call for tender No. 1409 of 14/09/2022 of Italian Ministry of University and Research funded by the European Union – NextGenerationEU Award Number: P2022RKXH9, Concession Decree No. 1397 of 06/09/2023 adopted by the Italian Ministry of University and Research, project CORonal mass ejection, solar eNERgetic particle and flare forecaSTing from phOtospheric sigNaturEs (CORNERSTONE).
\endgroup

\section{List of Acronyms}
\begin{acronym}[SWATNet Project]\itemsep2pt
\acro{1D}{1-Dimensional}
\acro{2D}{2-Dimensional}
\acro{3D}{3-Dimensional}
\acro{AA}{Academy of Athens}
\acro{AC}{Activity-Change}
\acro{AIA}{Atmospheric Imaging Assembly}
\acro{AI}{Artificial Intelligence}
\acro{AUC}{Area Under the Curve}
\acro{AR}{Active Region}
\acro{ASCII}{American Standard Code for Information Interchange}
\acro{ASRO}{Aboa Space Research Oy}
\acro{AU}{Astronomical Unit}
\acro{CAM}{Class Activation Maps}
\acro{CDAWeb}{Coordinated Data Analysis Web}
\acro{CDF}{Common Data Format}
\acro{CDP}{Career Development Plan}
\acro{CCD}{Charge Coupled Device}
\acro{CET}{Central European Times}
\acro{CI}{Confidence Interval}
\acro{CME}{Coronal Mass Ejection}
\acro{CNN}{Convolutional Neural Network}
\acro{CSV}{Comma-Seperated Values}
\acro{CV}{Cross-Validation}
\acro{DDIM}{Denoising Diffusion Implicit Model}
\acro{DDPM}{Denoising Diffusion Probabilistic Model}
\acro{DHO}{Debrecen Heliophysical Observatory}
\acro{DN}{Digital Nivel}
\acro{DTW}{Dynamic Time Warping}
\acro{DOI}{Digital Object Identifier}
\acro{EAB}{External Advisory Board}
\acro{ECAS}{European Commission Authentication System}
\acro{ECTS}{The European Credit Transfer and Accumulation System}
\acro{ELTE}{Eötvös Loránd University}
\acro{ESA}{European Space Agency}
\acro{ESR}{Early Stage Researcher}
\acro{EU}{European Union}
\acro{EUHFORIA}{EUropean Heliospheric FORecasting Information Asset}
\acro{EUHFORIA 2.0}{EUropean Heliospheric FORecasting Information Asset 2.0}
\acro{EUV}{Extreme Ultraviolet}
\acro{FARe}{False Alarm Rate}
\acro{FAR}{False Alarm Ratio}
\acro{FGE}{Fluid Gravity}
\acro{FITS}{Flexible Image Transport System}
\acro{FP7}{The Seventh Framework Programme of the European Union}
\acro{FN}{False Negative}
\acro{FP}{False Positive}
\acro{FPR}{False Positive Rate}
\acro{FSS}{F1-Skill-Score}
\acro{GA}{Grant Agreement}
\acro{GAN}{Generative Adversarial Network}
\acro{GCS}{Graduated Cylindrical Shell}
\acro{GSO}{Gyula Bay Zoltán Solar Observatory}
\acro{HEEQ}{HEliocentric Earth equatorial}
\acro{HMI}{Helioseismic and Magnetic Imager}
\acro{HSPF}{Hungarian Solar Physics Foundation}
\acro{HSS}{Heidke Skill Score}
\acro{ICME}{Interplanetary Coronal Mass Ejection}
\acro{IDL}{Interactive Data Language}
\acro{I/O}{Input/Output}
\acro{IMF}{Interplanetary Magnetic Field}
\acro{IP}{InterPlanetary}
\acro{IPN}{Instituto Pedro Nunes}
\acro{ITN}{Innovative Training Network}
\acro{JSOC}{Joint Science Operations Center}
\acro{KUL}{KU Leuven}
\acro{L1}{first Lagrangian point}
\acro{LOS}{Line-Of-Sight}
\acro{LPIPS}{Learned Perceptual Image Patch Similarity}
\acro{LSTM}{Long Short-Term Memory}
\acro{MAPE}{Mean Absolute Percentage Error}
\acro{MAE}{Mean Absolute Error}
\acro{MCC}{Matthews Correlation Coefficient}
\acro{MFM}{magnetofrictional method}
\acro{MHD}{magnetohydrodynamics}
\acro{ML}{Machine Learning}
\acro{MLP}{Multi-Layer Perceptron}
\acro{MPF}{Maximum Peak Flux}
\acro{MSCA}{Marie-Sk{\l}odowska-Curie Action}
\acro{NC}{No-Change}
\acro{NN}{Neural Network}
\acro{NFB}{Negative-Frequency-Bias}
\acro{NPV}{Negative Predictive Value}
\acro{NRT}{Near-Real-Time}
\acro{OA}{Open Access}
\acro{PCDP}{Personal Career Development Plan}
\acro{PFSS}{Potential Field Source Surface}
\acro{PHI}{Polarimetric and Helioseismic Imager}
\acro{PI}{Principal Investigator}
\acro{PIL}{Polarity Inversion Line}
\acro{PR-F1}{Persistent-Relative-F1}
\acro{PRSS}{Persistent Relative Skill Score}
\acro{PSNR}{Peak Signal To Noise Ratio}
\acro{PSP}{Parker Solar Probe}
\acro{PTECH}{Present Technologies, LDA}
\acro{RNN}{Recurrent Neural Network}
\acro{ROC}{Receiver Operating Characteristic}
\acro{RGB}{Red-Green-Blue}
\acro{SAS}{Space Applications Services}
\acro{SB}{Supervisory Board}
\acro{SDO}{Solar Dynamic Observatory}
\acro{SEP}{Solar Energetic Particle}
\acro{SF}{Solution Focus}
\acro{SOHO}{Solar and Heliospheric Observatory}
\acro{SolO}{Solar Orbiter}
\acro{SOLPACS}{SOLar Particle Acceleration in Coronal Shocks}
\acro{SPCNN}{Solar-Patch-Distributed-CNN}
\acro{SSC}{Sheffield Solar Catalogue}
\acro{SSIM}{Structural Similarity Index Measure}
\acro{STEM}{Science, Technology, Engineering and Mathematics}
\acro{STEREO}{Solar Terrestrial Relations Observatory}
\acro{SVM}{Support Vector Machine}
\acro{SWATNet}{Space Weather Awareness Training Network}
\acro{SXR}{Soft X-Rays}
\acro{T2PF}{Time To Peak Flux}
\acro{TN}{True Negative}
\acro{TNR}{True Negative Rate}
\acro{TP}{True Positive}
\acro{TPR}{True Positive Rate}
\acro{TSS}{True Skill Statistic}
\acro{UC}{University of Coimbra}
\acro{UH}{University of Helsinki}
\acro{UMCS}{Maria Curie-Skłodowska University}
\acro{UNITOV}{Università degli Studi di Roma Tor Vergata}
\acro{UoI}{University of Ioannina}
\acro{USFD}{University of Sheffield}
\acro{UTU}{University of Turku}
\acro{UV}{Ultraviolet}
\acro{VAE}{Variational Auto Encoder}
\acro{VideoLENS}{Video Local Event Neural System}
\acro{WP}{Work Package}

\end{acronym}

\bibliographystyle{apalike-fr}
\bibliography{bib.bib} 

\end{document}